\documentclass{PoS}
\usepackage{latexsym}
\usepackage{amsmath}
\usepackage{amssymb}
\usepackage{graphicx}
\usepackage{colordvi}
\usepackage{multirow}
\usepackage{epsfig,graphics}

\title{Stable and quasi-stable confining $SU(N)$ strings in $D=2+1$.}

\ShortTitle{Stable and Quasi-Stable SU(N) flux-tubes.}

\author{\speaker{Andreas Athenodorou} \\
        Swansea University, Singleton Park, Swansea, SA2 9GX, UK \& \\
        University of Cyprus, Nicosia, CY-1678, Cyprus \\ 
        E-mail: \email{athenodorou.andreas@ucy.ac.cy}}

\author{Michael Teper \\
        University of Oxford, Rudolf Peierls Center for Theoretical Physics, 1 Keble Road, OX1 3NP, Oxford, UK\\
        E-mail: \email{m.teper1@physics.ox.ac.uk}}

\abstract{We investigate the low-lying spectrum of closed confining flux-tubes that wind around a spatial torus in $D=2+1$ and carry
flux in different representations of $SU(N)$. We focus on our most recent calculations for $N=6$ and $\beta = 171$, where 
the calculated low-energy physics is very close to the continuum and large-$N$ limits.
We investigate the adjoint, {\bf{\underline{84}}}, {\bf{\underline{120}}}, $k=2A$, $2S$ and $k=3A, \ 3M, \ 3S$ representations
and show that the corresponding flux-tubes do exist. Similarly to the results for the fundamental 
representation, the ground state of a flux-tube with momentum along its axis appears
to be well described by Nambu-Goto all the way down to very short tubes. 
In contrast, excited states have much larger deviations from Nambu-Goto. We discuss whether these states are non-string-like and associated 
with excitations of massive flux-tube modes.}

\FullConference{The XXXI International Symposium on Lattice Field Theory\\
         July 29 - August 03 2013\\
         Mainz, Germany}

\begin{document}
\vskip -0.5125cm
\section{Introduction}
\label{Introduction}
\vskip -0.5125cm
In 2+1 dimensional $SU(N)$ gauge theories and in the confining phase, the flux-tube which joins colour sources in the fundamental representation for large separations $l$ looks like a thin string. The spectrum of such a long flux-tube should be calculable in the framework of an effective string action. Recently, remarkable progress in determining the universal terms
of this effective string action~\cite{Ofer1} has shown that the large-$l$ spectrum is close to the Nambu-Goto `free string' spectrum. At the same time, and more surprisingly, numerical lattice calculations~\cite{ABM1,ABM2,ABM3} have shown that the spectrum of the confining flux-tube at small to medium values of $l$ is also remarkably close to Nambu-Goto. The very recent and remarkable work in \cite{SDRFVG_13} has begun to understand this is terms of the approximate integrability of the inter-phonon interactions.
 
In this work we extend our recent calculation~\cite{ABM1} of the fundamental flux-tube spectrum, $N=6$ and $\beta=171$ to colour representations higher than the fundamental. More specifically we investigate cases where the flux-tube is stable for all lengths and cases where it is not; these include representations with $N$-ality $k=0,1,2,3$. As before we focus on $SU(6)$, where the theory is close to the large-$N$ limit for a number of low-energy quantities. We have performed our calculations at a fixed lattice spacing which is small enough for many of the corrections to be negligible. Our goal in this work is to investigate whether Nambu-Goto provides a good description for string-like flux-tube states in colour representations higher than the fundamental and whether these spectra include supplementary non-string-like states associated to excitations of massive flux-tube modes. 
 
We begin our investigation with flux-tubes which would emanate from charges in the $k=2$ antisymmetric ($2A$) and symmetric ($2S$) irreducible representations arising from the tensor product of two fundamental colour charges. Subsequently, we move to the $k=3$ antisymmetric ($3A$), mixed ($3M$) and symmetric ($3S$) irreducible representations arising from the tensor product of three fundamental colour sources. Next we consider the $k=0$ adjoint flux-tube emanating from the product of a fundamental and an antifundamental source. Finally, we probe the $k=1$ representations {\underline {\bf 84}}, {\underline {\bf 120}} arising from the product of two fundamental and an antifundamental source.    
\vspace{-0.5125cm}
\section{Expectations}
\label{expectations}
\vspace{-0.5125cm}
We focus on the spectrum of closed flux-tubes that wind around a spatial torus of length $l$. In order to avoid any finite volume corrections we make the transverse and temporal tori, $l_{\perp}$ and $l_t$ respectively, large enough. As the flux-tube gets shortened at some point one confronts the finite volume transition at length $l_c=1/T_c$ ($T_c$ the deconfining temperature) below which the flux-tube dissolves. Since $T_c$ is of the order of $\sqrt{\sigma_f}$ our flux-tubes have lenghts $l > 1 / \sqrt{\sigma_f}$.

The flux-tubes we will study have non-zero longitudinal momentum winding along the flux with value $2 \pi q/l$ and $q=0,1,2$. One can think that in the large-$N$ limit and, thus, in absence of handles and branchings the worldsheet has a topology of a cylinder. The simplest action one can consider, in this case, is just the area of the world sheet swept by the propagation of the string in a flat space-time; in other words the Nambu-Goto action~\cite{Arvis}. This model is self-consistent quantum mechanically only in $D=26$ dimensions but it is expected that, for any arbitrary $D$, this model can be used as an effective low energy field theory for long enough strings. The Nambu-Goto energy spectrum for the closed string in $D=3$ is given by the following expression:
\begin{equation}
E_{N_L, N_R}(q,l)
=
\left\{
(\sigma_{\cal R} l)^2 
+
8\pi\sigma_{\cal R} \left(\frac{N_L+N_R}{2}-\frac{1}{24}\right)
+
\left(\frac{2\pi q}{l}\right)^2\right\}^{\frac{1}{2}}
\label{eqn_EnNG}
\end{equation}
Here $\sigma_{\cal R}$ is the string tension at colour representation ${\cal R}$. The energy contribution of the mass-less phonons propagating along the string is encoded in $N_L = \sum_k  n_L(k)k$ and $N_R = \sum_k  n_R(k)k$ where $n_{L(R)}(k)$ is the number of left(right) moving phonons of momentum $|p|=2\pi k/l$. The difference of the occupation numbers $N_L$ and $N_R$ is related to the longitudinal momentum through the level matching constraint $N_L - N_R = q$. Finally, the parity of a state is given by $P = (-1)^{number\ of\ phonons}$.
\vspace{-1.1525cm}
\section{Lattice Calculation}
\label{latticeformulation}
\vspace{-0.5525cm}
Our gauge theory is defined on a three-dimensional Euclidean space-time lattice that has been toroidally compactified with $L_{x} \times L_{\perp} \times L_T$ sites with $l=aL_x$; $l_{\perp}=aL_{\perp}$ and $l_{T}=aL_T$. For the spectrum calculation we perform Monte-Carlo simulations using the standard Wilson plaquette action
$S_{\rm W}=\beta \sum_P \left[ 1- \frac{1}{N} {\rm Re} {\rm Tr}{U_P} \right]$. 
The bare coupling $\beta$ is related to the dimensionful coupling $g^2$ through $\lim_{a\to 0}\beta={2N}/{ag^2}$. In the large--$N$ limit, the 't Hooft coupling $\lambda=g^2N$
is kept fixed. 
 The simulation we use combines standard heat-bath and over-relaxation steps in the
ratio 1 : 4. These are implemented by updating $SU(2)$ subgroups using the Cabibbo-Marinari algorithm.

We calculate energies from the time behaviour of correlators of
suitable operators $\{\phi_i\}$, such as
$C_{ij}(t) = \langle \phi_i^\dagger(t)\phi_j(0) \rangle = \langle \phi_i^\dagger e^{-Han_t} \phi_j \rangle = \sum_k c_{ik} c^\star_{jk} e^{-aE_k n_t}.$
We project onto loops of flux closed around the $x$-torus. Hence, we use operators that
wind around the $x$-torus. The simplest such operator is the Polyakov loop $\phi(n_y,n_t) = \mathrm{Tr}_{\cal R} \{l_p(n_y,n_t)\}$ with $l_p(n_y,n_t) = \prod^{L_x}_{n_x=1} U_x(n_x,n_y,n_t)$. Here we have taken the product of
the link matrices in the $x$-direction, around the $x$-torus and the trace is taken in the
desired representation $\cal R$. Carrying out the tensor product decomposition for each irreducible representation ${\cal R}$ we obtain that $\mathrm{Tr}_{\rm Adj}\{ l_p  \} =  \mathrm{Tr} \{ l_p \}  \mathrm{Tr} \{ l_p^{\dagger} \} - 1 $,  $\mathrm{Tr}_{f}\{ l_p  \} =  \mathrm{Tr} \{ l_p \}$,  $ \mathrm{Tr}_{{\underline {\bf 84}}({\underline {\bf 120}})} \{ l_p \}= \frac{1}{2} \left[\{{\mathrm{Tr}} \{ l_p \}\}^2 -(+) {\mathrm{Tr}} \{ l^2_p \}   \right] {\mathrm{Tr}} \{  l^{\dagger}_p \}  -{\mathrm{Tr}} \{ l_p \}$, \ \ $\mathrm{Tr}_{{2A}({2S})} \{ l_p \} = \frac{1}{2} \left[  \left\{ {\mathrm{Tr}} \{ l_p \} \right\}^2 -(+) {\mathrm{Tr}} \{ l^2_p \}  \right]$, \ \ $\mathrm{Tr}_{{3A}({3S})} \{ l_p \} = \frac{1}{6} \left[  \left\{ {\rm Tr} \{ l_p \} \right\}^3 - (+) 3 {\rm Tr} \{ l_p \}  {\rm Tr} \{ l^2_p \}  + 2 {\rm Tr} \{ l^3_p \}  \right]$  \ and \ $\mathrm{Tr}_{{3M}} \{ l_p \} = \frac{1}{3} \left[  \left\{ {\rm Tr} \{ l_p \} \right\}^3 - {\rm Tr} \{ l^3_p \}  \right]$. In addition we also use many other winding paths, as listed in Table 2 of \cite{ABM1}, and also we smeared and blocked the $SU(N)$ link matrices. Using all these paths we project onto different longitudinal momenta and parities keeping the transverse momentum equal to $p_\perp=0$. Subsequently we perform a variational calculation of the spectrum, maximising $ \langle e^{-Ht} \rangle$  over this basis. This provides us with an ordered set of approximate energy eigenoperators  $\{\psi_i\}$. We then form the correlators of these, $\langle \psi^\dagger_i(t) \psi_i(0) \rangle$, and extract the energies from plateaux in the effective energies.
\vspace{-0.65cm}
\section{Results}
\label{Results}
\vspace{-0.35cm}
\subsection{Fundamental Representation}
\label{fundmanental_Representation}
\vspace{-0.30cm}
In our previous work~\cite{ABM1} we performed calculations in $SU(6)$ for the closed flux-tube spectrum on the same lattices,
and at the same coupling as for this work. There we observed that the absolute ground state is very accurately described by the free string
prediction in Equation~(\ref{eqn_EnNG}), with a correction only becoming visible for $l\surd\sigma_f \lesssim 2$. The lightest states with $q \neq 0$ also showed no visible correction to Nambu-Goto down to  $l\surd\sigma_f \sim 1.5$. While other low-lying excited states typically show larger corrections, these 
typically become insignificant at values of $l$ that are much smaller than required for the expansion of Equation~(\ref{eqn_EnNG}) in powers of $1/l^2\sigma$ 
to become convergent. Finally, we see no evidence for any non-stringy massive modes.
\vspace{-0.35cm}
\subsection{$k=2$ Antisymmetric and Symmetric Representations}
\label{k2}
\vspace{-0.30cm}
In the $k=2$ sector we focus on the totally antisymmetric, 2A and the totally symmetric, 2S representations. The lightest $k = 2$ flux-tube is pure $k = 2A$ and it is lighter than two fundamental flux-tubes; thus it is stable. In Figure~\ref{figure1}(a) we provide the results for the $q=0,1,2$ ground states. The pure Nambu-Goto prediction appears to fit very well these states. For the absolute ground state ($q=0$) the expansion of the Nambu-Goto prediction for the energy  $E_{0,0}(0,l)$ in powers of $1/l^2\sigma$ converges right through the range of $l$ where we have calculations all the way down to  $l\surd\sigma = \pi/3 \sim 1.1 < l_c\surd\sigma$. We see that the free string expression is good all the way down to $l\surd\sigma_{2A} \sim 2$ which is close to the deconfining length. In addition a  $O(1/l^7)$ correction can describe the deviations from Nambu-Goto for $l\surd\sigma_{2A} \leq 2$.  

On the other hand the spectrum of $2A$ excited states with $q=0$ appears to suffer from large deviations from the Nambu-Goto free-string expectations. We plot, in Figure~\ref{figure1}(b), the four lightest $P=+$ states, and the two lightest $P=-$ ones, as well as the predictions of Nambu-Goto for the lowest few energy levels. In Nambu-Goto the first excited state, like the ground state, is non-degenerate with one left and one
right moving phonon with momenta $p = \pm 2 \pi /l$ while the next energy level has four degenerate
states with the left and right moving phonons sharing twice the minimum momentum. This can be carried by one or two phonons, thus, two of these states have $P=+$ and two have $P= -$. The first excited state has large deviations from the Nambu-Goto curve but appears to approach the string prediction. This means that it is either asymptoting to that curve or it is just crossing it. For the next excited level, the observed excited states are far from showing the Nambu-Goto degeneracies and far from the Nambu-Goto predicted energies, even for the largest values of $l \sqrt{\sigma}$. These large discrepancies raise the question whether these states are string like or associated with some massive modes; this is answered in the next section. 
 
We turn now to the $2S$ representation. The associated states are expected to be heavier
than the corresponding $2A$ states and, thus, unstable with larger statistical errors. In Figure~\ref{figure1}(c) we show the ground states with $q=0,1,2$. Just as for $k=2A$ the energies are remarkably close to the Nambu-Goto predictions. Although unstable, the absolute ground state possess extended plateaux very different from the decay thresholds. In contrast to $k=2A$, the situation with the excited $q=0$ states is however much worse and we are unable to extract the corresponding energies.
\vspace{-0.45cm}
\subsection{$k=3$ Antisymmetric, Mixed and Symmetric Representations}
\label{k3}
\vspace{-0.30cm}
In the $k=3$ sector we focus on the totally antisymmetric, $3A$, mixed, $3M$, and totally symmetric, $3S$ representations. Their string tensions are known from previous work to be close to the Casimir scaling, thus, we know that $k=3A$ is stable, $3M$ nearly stable and $3S$ highly unstable. 

In Figure~\ref{figure2}(a) we plot the lightest energies of $k=3A$ flux-tubes with $q=0,1,2$. Just as for the corresponding $k=2$ flux-tubes, we see excellent agreement
with Nambu-Goto all the way down to $l \sim l_c$. The relevant asymptotic decay states are heavier than the $k=3A$ states demonstrating, thus, their stability. Turning now to the excited states for $q=0$ we observe that their qualitatively behaviour resembles that for $k=2A$. Namely, the excited states suffer with large deviations from Nambu-Goto predictions, raising the question whether these states are string like or associated with excitations of massive flux-tube modes.   
 
We focus now on the heavier $k=3M$ states. We plot in Figure~\ref{figure2}(b) the ground states with $q=0,1,2$. Once again these particular states agree very well with Nambu-Goto. We also observe that their instability is not enough to affect the extraction of the states using effective mass plateaux. However, the $q=0$ excited states are very unstable and we are unable to identify useful plateaux.

The $k=3S$ states are much heavier and we can only estimate energies for the $q=0,1$ ground states; these are presented in Figure~\ref{figure2}(c). Again we see rough agreement with Nambu-Goto.
\vspace{-0.35cm}
\subsection{$k=0$ Adjoint Representation}
\label{k0}
\vspace{-0.30cm}
Old calculations~\cite{Bali} provide some evidence that such flux-tubes exist with string tension approximately proportional to the Casimir Scaling. Adjoint flux-tubes can be screened down to the vacuum by gluons but this is suppressed by $1/N^2$. Hence one expects to observe such flux-tubes at large enough $N$. An adjoint flux-tube is in general heavier than a pair of fundamental and anti-fundamental flux-tubes and can decay into these (for not so large $N$ and $l$). 

In Figure~\ref{figure3}(a) the energies of the lightest adjoint flux-tubes with $q=0,1,2$ show that Nambu-Goto provides a remarkable description all the way down to $l\sim l_c$. The decay thresholds are indicated (dashed lines) and we see that the decay phase space is small, raising the hope that the decay widths will be negligibly small. Concerning the excitation spectrum, we could not identify well-defined excited states with $q=0$. This is somehow reasonable since these states would have a very large phase space for decay into a pair of fundamental and antifundamental flux-tubes.
\vspace{-0.45cm}
\subsection{$k=1$ {\underline {\bf 84}} and {\underline {\bf 120}} Representations}
\label{k1}
\vspace{-0.30cm}
Here we study flux-tubes carrying flux in the $k=1$ {\underline {\bf 84}} and {\underline {\bf 120}} representations. Such flux-tubes can mix with single fundamental flux-tubes, but this is large-$N$ suppressed and we, thus, ignore this possibility. Nevertheless, the decay/mixing with 3 (anti)fundamental flux-tubes and with a $k=2A$ and an antifundamental flux-tube is not large-$N$ suppressed. For {\underline {\bf 84}} in Figure~\ref{figure3}(b) we observe that the agreement with Nambu-Goto is, once again, remarkably good for $q = 0,1$ and quite good for $q = 2$. The excited states are very massive and it becomes difficult to identify plausible plateaux. 

In Figure~\ref{figure3}(c) we plot the ground state energies of flux-tubes in the {\underline {\bf 120}} representation for $q = 0, 1$. Once again we observe that Nambu-Goto provides an adequate description. The  {\underline {\bf 120}} string tension is large, prohibiting us from extracting energy mass plateaux for $q=2$ and for excitations with $q=0$. 
\begin{figure}
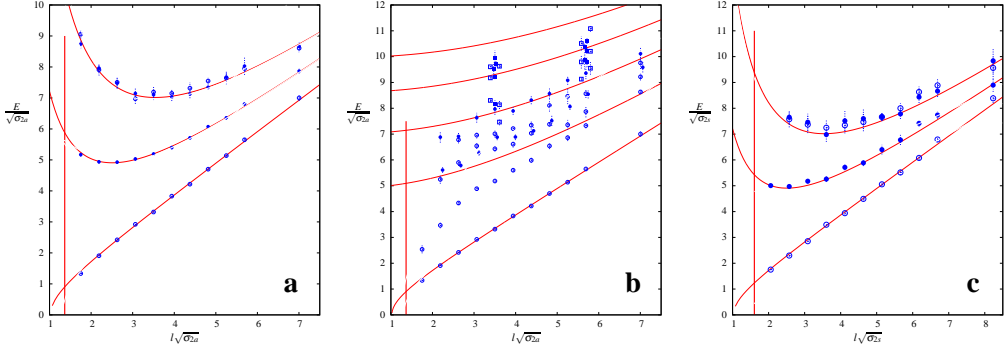

\vspace{-0.65cm}
\begin{center}
{\scalebox{0.35}{\input{plot_EgsQallk2a_n6fb.tex}}\put(-20,20){\bf a}}
{\scalebox{0.35}{\input{plot_Eq0k2a_n6f.tex}}\put(-20,20){\bf b}}
{\scalebox{0.35}{\input{plot_EgsQallk2s_n6f.tex}}\put(-20,20){\bf c}}
\end{center}
\vspace{-0.5cm}
\caption{({\bf a}): $k=2A$ ground states with $q=0,1,2$. ({\bf b}) Energies of lightest $k=2A$ $q=0$ states. ({\bf c}) $k=2S$ ground states with $q=0,1,2$. $P=+$ states are represented by \Blue{$\circ$}, and  $P=-$, by \Blue{$\bullet$}. Solid red curves are Nambu-Goto predictions. Vertical line denotes location of `deconfinement' transition.}
\label{figure1}
\vspace{-0.45cm}
\end{figure}
\begin{figure}
\vspace{-0.65cm}
\begin{center}
{\scalebox{0.35}{\input{plot_EQgsk3A_n6f.tex}}\put(-20,20){\bf a}}
{\scalebox{0.35}{\input{plot_EQgsk3M_n6f.tex}}\put(-20,20){\bf b}}
{\scalebox{0.35}{\input{plot_EQgsk3S_n6f.tex}}\put(-20,20){\bf c}}
\end{center}
\vspace{-0.45cm}
\caption{({\bf a}): $k=3A$ ground states with $q=0,1,2$. ({\bf b}) $k=3M$ ground states with $q=0,1,2$. ({\bf c}) $k=3S$ ground states with $q=0,1$. $P=+$ states are represented by \Blue{$\circ$}, and  $P=-$, by \Blue{$\bullet$}. Solid red curves are Nambu-Goto predictions. Vertical line denotes location of `deconfinement' transition.}
\vspace{-0.35cm}
\label{figure2}
\end{figure}

\begin{figure}
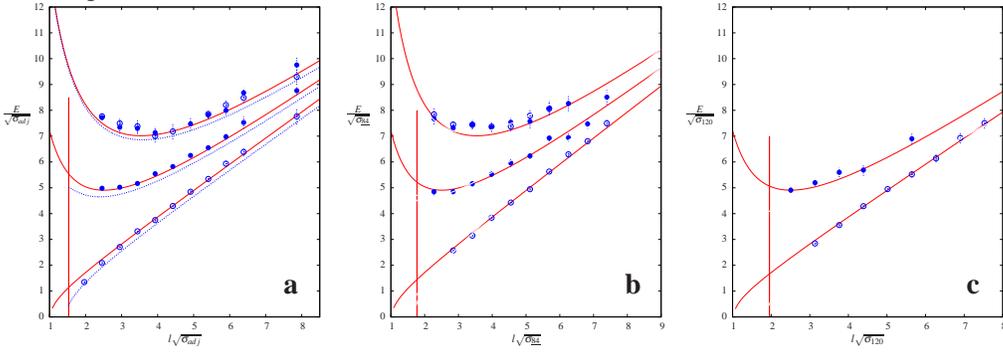

\vspace{-0.15cm}
\begin{center}
{\scalebox{0.35}{\input{plot_EgsQalladj_n6f.tex}}\put(-20,20){\bf a}}
{\scalebox{0.35}{\input{plot_EgsQallr84_n6f.tex}}\put(-20,20){\bf b}}
{\scalebox{0.35}{\input{plot_EgsQallr120_n6f.tex}}\put(-20,20){\bf c}}
\end{center}
\vspace{-0.5cm}
\caption{({\bf a}): Adjoint ground states with $q=0,1,2$. ({\bf b}) {\underline{\bf 84}} ground states with $q=0,1,2$. ({\bf c}) {\underline{\bf 120}} ground states with $q=0,1$. $P=+$ states are represented by \Blue{$\circ$}, and  $P=-$, by \Blue{$\bullet$}. Solid red curves are Nambu-Goto predictions. Vertical line denotes location of `deconfinement' transition.}
\label{figure3}
\vspace{-0.55cm}
\end{figure}

\vspace{-0.65cm}
\section{Stringy or Massive? - Heuristic investigation}
\label{Stringy_or_massive}
\vspace{-0.35cm}
We expect bound states of fundamental flux-tubes to have a low-lying excitation spectrum that contains clear signatures of the binding scale. Our cleanest spectra in this paper are for $k=2A$ and $k=3A$ so we shall focus on these. So does the $k=2A(3A)$ $q=0$ spectrum shown in Figure~\ref{figure1}(b) reveal any massive modes that are additional to the stringy excitations which, at large $l$, tend to the Nambu-Goto curves? Since the low-lying excitation spectrum of fundamental flux tubes appears to contain only stringy states, it is interesting to compare it with our $k=2A$ spectrum. The immediate question this comparison raises is whether the first excited $k=2A$ state might be a massive mode with the second excited state being the first excited stringy mode. If this state is an approximate Nambu-Goto-like string excitation then we would expect its wave-functional to be similar to that of the first excited $k=1$ state,
which we have good reason to think of as being stringy.

\begin{figure}
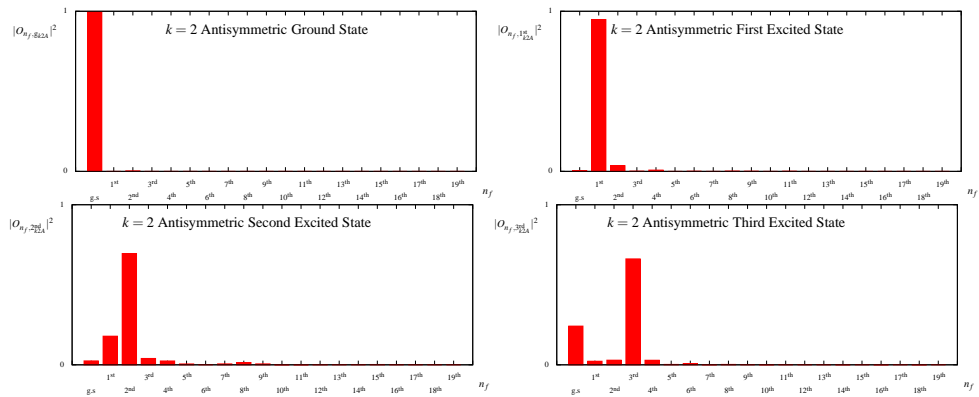

\vspace{-1.45cm}
\begin{center}
{\scalebox{0.50}{\input{plot_heuristic_ground_K2AS.tex}\put(-250,180){$k=2$ Antisymmetric Ground State}}}
{\scalebox{0.50}{\input{plot_heuristic_first_K2AS.tex}\put(-280,180){$k=2$ Antisymmetric First Excited State}}}
\end{center}
\vspace{-2.75cm}
\begin{center}
{\scalebox{0.50}{\input{plot_heuristic_second_K2AS.tex}\put(-280,180){$k=2$ Antisymmetric Second Excited State}}}
{\scalebox{0.50}{\input{plot_heuristic_third_K2AS.tex}\put(-280,180){$k=2$ Antisymmetric Third Excited State}}}
\vspace{-1.35cm}
\end{center}
\caption{Overlap, of $k = 2A$ ground, first, second and third excited $q=0$ states onto
the 20 lowest-lying fundamental states for $l = 52a$}
\label{figure4}
\vspace{-0.35cm}
\end{figure}

The comparison is made as follows. Let $\{\phi_i; i=1,...,n_o\}$ be our set of winding flux tube operators. When we perform our variational calculation over a basis in representation ${\cal R}$, we obtain a set of wavefunctionals, $\Phi^{n}_{\cal R}$ which can be written as linear combinations of our basis operators ${\rm Tr}_{\cal R} (\phi_i)$ such as $
\Phi^{n}_{\cal R}
=
\sum^{n_o}_{i} b^{n}_{{\cal R},i} c_{{\cal R},i}{\rm Tr}_{\cal R} (\phi_i)
\equiv
\sum^{n_o}_{i} b^{n}_{{\cal R},i} {\rm Tr}^{\prime}_{\cal R} (\phi_i)
$.
The coefficients $c_{{\cal R},i}$ have been chosen so that they satisfy the normalisation condition: 
$
\langle {\rm Tr}^{\prime\dagger}_{\cal R} (\phi_i(0))
{\rm Tr}^{\prime}_{\cal R} (\phi_i(0))\rangle = 1
$. This is to ensure that the comparison between the coefficients $b^{n}_{{\cal R},i}$ , which encode the ``shape'' of the state corresponding to the wavefunctional, of two different representations ${\cal R}$ can be meaningful. We, therefore, make the substitution 
${{\Phi}}^n_{\cal R}
=
\sum^{n_o}_{i} b^{n}_{{\cal R},i} {\rm Tr}^{\prime}_{\cal R} (\phi_i)
\to 
{\tilde{\Phi}}^n_{\cal R}
=
\sum^{n_o}_{i} b^{n}_{{\cal R},i} {\rm Tr}^{\prime}_{f} (\phi_i)
$
in order to compare an excited or ground state $k=1$ and ${\cal R}=2A$ wavefunctionals by comparing ${\tilde{\Phi}}^n_{\cal R}$ with ${{\Phi}}^n_{f}$ for the fundamental representation. This can be quantified by calculating the overlap
$O_{n^\prime,n}({\cal R}=2A)=
\frac{ \langle {\Phi^{n^\prime\dagger}}_f {\tilde{\Phi}^n}_{\cal R} \rangle}
{\langle{\Phi^{n^\prime\dagger}}_f \Phi^{n^\prime}_f \rangle^{1/2} 
\langle {{\tilde{\Phi}}}^{n\dagger}_{\cal R} \tilde{\Phi}^{n}_{\cal R} \rangle^{1/2} }$.
Examples of these overlaps are presented in Figure~\ref{figure4}. These provide us with a measure of the similarity between 
the original state ${{\Phi}}^n_{2A}$ and the state ${{\Phi}}^{n^\prime}_{f}$ for $q=0$.  For the ground state we observe an overwhelming similarity ($\sim 100 \%$) between the associated wavefunctionals for $k=2A$ and $f$. Turning now on the first excited state $k=2A$ we find that by far the largest overlap is indeed
on the first excited fundamental with a very small overlap onto the $2^{\rm nd}$ excited stringy $f$ state. Hence, this state is definitely not some new massive mode excitation. Concerning the next two $k=2A$ excitations, while the dominant overlap
is onto the corresponding fundamental excited state, there is a large projection on other states as well. It certainly appears possible that
some new massive mode either dominates or is mixed into one or both of these states. Similar results have been obtained for $k=3A$. 

\vspace{-0.45cm}
\section{Conclusion}
\label{conclusion}
\vspace{-0.45cm}
We have calculated the low-lying spectrum of closed flux-tubes in various representations of colour. One of the aims of this project was to compare the resulting spectrum to simple effective string actions, just as we did in our previous work on flux-tubes in the fundamental representation of colour. Flux-tubes in irreducible representations higher than the fundamental can be thought of as bound states of fundamental and antifundamental flux-tubes. Hence the massive excitation modes related to that binding might leave their signature in the spectrum. While in the case of fundamental flux-tubes we found no sign of such states we hoped that we would find something different here. 

Our results show clearly that the absolute ground state and the lightest states with non-zero longitudinal momenta are accurately described by the free string expression shown in Equation~(\ref{eqn_EnNG}). Our results also suggest that the lightest two $p=0$ states of the higher than the fundamental flux-tubes consist only of string-like states. For a more detailed description of this project you can look at the longer write-up~\cite{AA4}.
\vspace{-0.45cm} 
\section*{Acknowledgements}
\label{Acknowled}
\vspace{-0.45cm}
This work is a continuation of a project in the initial steps of which Barak Bringoltz, whom we thank, has played an essential role in developing many of the ideas and techniques used in this work. The computations were carried out on computational resources in Oxford Theoretical Physics funded by EPSRC and Oxford University. AA would like to thank Swansea University and especially Biagio Lucini for their support.

\vspace{-0.225cm}
\end{document}